\documentclass[aps,prl,amsmath,amssyb,floatfix,twocolumn,10pt]{revtex4-1}
\usepackage{graphicx}
\usepackage{hyperref}
\usepackage{braket}
\usepackage{bm}

\def\prl{{Phys. Rev. Lett. }}
\def\prb{{Phys. Rev. B }}

\def\rmp{{Rev. Mod. Phys. }}
\def\science{{Science }}
\def\nature{{Nature }}
\def\natphys{{Nat. Phys. }}
\def\natcomm{{Nat. Commun. }}

\def\ie{{\it i.e.}}

\def\bPhi{{\bm \Phi}}
\def\bnabla{\bm{\nabla}}
\def\rhos{\rho_s}
\def\D{\mathcal{D}}

\def\br{{\bf r}}
\def\bu{{\bf u}}

\def\bk{{\bf k}}
\def\bq{{\bf q}}
\def\bQ{{\bf Q}}
\def\bA{{\bf A}}

\def\bR{{\bf R}}
\def\bK{{\bf K}}
\def\bQ{{\bf Q}}

\def\dr{{d^2r}}
\def\Tr{{\rm Tr}}

\def\YBCO{YBa$_2$Cu$_3$O$_{6+x}$}
\def\LSCO{La$_{2-x}$Sr$_x$CuO$_4$}
\def\HBCO{HgBa$_2$CuO$_{4+\delta}$}

\def\BSCCO{Bi$_2$Sr$_2$CaCu$_2$O$_{8+x}$}
\def\tU{\tilde{U}}
\def\tJ{\tilde{J}}

\begin{document}

\title{Dimensional Crossover of Charge-Density Wave Correlations in the Cuprates}

\author{Yosef~Caplan}
\affiliation{Racah Institute of Physics, The Hebrew University,
  Jerusalem 91904, Israel}
\author{Dror~Orgad}
\affiliation{Racah Institute of Physics, The Hebrew University,
  Jerusalem 91904, Israel}

\date{\today}

\begin{abstract}

Short-range charge-density wave correlations are ubiquitous in underdoped cuprates. They are largely confined
to the copper-oxygen planes and typically oscillate out of phase from one unit cell to the next in the $c$-direction.
Recently, it was found that a considerably longer-range charge-density wave order develops in \YBCO~above a sharply
defined crossover magnetic field. This order is more three-dimensional and is in-phase along the $c$-axis.
Here, we show that such behavior is a consequence of 
the conflicting ordering tendencies
induced by the disorder potential and the Coulomb interaction, where the magnetic field acts to tip the scales
from the former to the latter. We base our conclusion on analytic large-$N$ analysis and Monte-Carlo simulations
of a non-linear sigma model of competing superconducting and charge-density wave orders. Our results are
in agreement with the observed phenomenology in the cuprates, and we discuss their implications
to other members of this family, which have not been measured yet at high magnetic fields.

\end{abstract}

\pacs{74.72.Kf,75.10.Hk,74.62.En,74.40.-n}

\maketitle

The cuprate high-temperature superconductors are in a delicate state of balance between
various electronic orders \cite{intertwined}. In particular, experiments have revealed a
subtle interplay between the superconducting (SC) and charge-density wave (CDW) orders.
Much of the evidence for the latter, coming from x-ray scattering
\cite{Ghiringhelli,Chang,Achkar1,Blackburn13,Achkar2, Comin1,daSilva,Le-Tacon,Croft,
Christensen,Tabis,Campi,SilvaNeto,Comin3,Comin-symmetry,Forgan-structure,Peng-Bi2201} and nuclear magnetic
resonance (NMR) measurements \cite{NMR-shortcor}, points at short-range, in-plane, CDW order
which is in competition with superconductivity. Concretely, the intensity of the CDW scattering
peak grows as the system is cooled towards the SC transition temperature, $T_c$, and then
decreases or saturates upon entering the SC phase. Furthermore, the CDW signal is enhanced when
a magnetic field is used to partially quench superconductivity.

However, recent x-ray scattering measurements of \YBCO $\,$(YBCO) \cite{Gerber3D,Chang3D,Jang3D} have
detected additional Bragg peaks that are different in several respects from the signal described above.
First, the new peaks are much sharper, thus corresponding to considerably longer-ranged
CDW correlations. Secondly, whereas both the short-range and long-range CDW peaks
share the same incommensurate in-plane wave-vector, the latter appear only along the $b$-direction
and at integer $c$-axis wave-vectors (measured in reciprocal lattice units), $l$. This stands in
contrast to the bidirectional nature and the half-integer $l$ of the former.
Thirdly, as also found by NMR \cite{NMR-nature,NMR-nature-comm} and ultrasound measurements \cite{ultrasound},
the longer-range CDW order sets in only above a magnetic field $H_{3D}\approx 15$T,
and at temperatures below $T_{3D}\approx 50$K. The short-range correlations, however,
appear already at zero field and survive up to about $T_{ch}=150$K.


Previously, aspects of competing SC and CDW orders were studied via Ginzburg-Landau
and non-linear sigma models (NLSM) \cite{Zachar,Demler,Efetov,Hayward1,Nie,NLSM-layers,Meier,Einenkel-vortex}.
Those directly related to recent experiments include the CDW temperature dependence \cite{Hayward1},
the effects of disorder \cite{Nie,NLSM-layers} and of a magnetic field \cite{NLSM-layers,Meier,Einenkel-vortex}.
However, a framework in which to understand the complete phenomenology, especially the
relation between short and long-range CDW order in YBCO, is lacking. Here, we offer such a scheme
by including in our recent NLSM \cite{NLSM-layers} the structure and couplings of YBCO,
by elucidating the different effects of the disorder on the chain layers and on the CuO$_2$ planes,
and by going beyond the inter-plane mean-field approximation. The gained insights are then applied to
other cuprates.

We use analytical large-$N$ and replica techniques alongside
Monte-Carlo simulations to show that the physics is driven by opposing forces.
While the Coulomb interaction causes the CDW order to change sign from one plane to the next within a
CuO$_2$ bilayer, its relative phase between consecutive unit cells in the $c$-direction is frustrated. On the
one hand, the disordered dopant potential on the chain layer tends to induce the same CDW configuration on the
two adjacent CuO$_2$ bilayers. On the other hand, such an arrangement is costly from the point
of view of their mutual capacitive energy, which is minimized by having them host out-of-phase CDWs.

At zero magnetic field the disorder prevails. CDW puddles that nucleate at locally favorable potential regions
on the nearest CuO$_2$ planes to the chain layer, tend to be in phase to each other and opposite to the CDW order
that develops on the other CuO$_2$ plane within their bilayer. This leads to a CDW structure factor that is
centered near half-integer $l$ with a $c$-axis correlation length, $\xi_c$, of about one lattice constant.
The in-plane correlation lengths, $\xi_{a,b}$, are longer but still extend over only few wave periods.
For disorder fluctuations that are larger than the slight anisotropy induced by the chains, which
benefits $b$-axis CDW, the nucleated CDW regions are distributed evenly between the $a$ and $b$
directions, thus leading to scattering peaks in both directions \cite{DelMaestro,Robertson}.

A magnetic field introduces vortices into the system at which superconductivity is suppressed and the CDW
amplitude is significantly larger than its typical value without the field \cite{Hamidian-nematic}.
This in turn implies that the
inter-layer Coulomb interaction and the chain-induced anisotropy play a more important role in the energetic
balance governing these regions. As an outcome, the CDW halos formed around a vortex line tend to order
with integer $l$ in the $c$-direction and orient preferentially, although not exclusively, along the $b$-axis.
The disorder, on its part, interferes with the establishment of inter-halo coherence, both along the vortex and
more importantly between different vortices. However, as the field is made stronger vortices move closer together
until correlations between the $b$-oriented halos start rapidly increasing. Our calculations indicate that
this growth would eventually turn into true long-range order at a critical field. Such a transition
is possible since the chain disorder couples to the gradient of the integer-$l$ CDW order, thereby
reducing the lower critical dimension to $d_L=2$. In contrast, any disorder on the CuO$_2$
layers, for which $d_L=4$ \cite{Imry-Ma}, would smear the transition into a crossover.
Nevertheless, as long as this disorder is not too strong the
high-field state will still exhibit unidirectional integer-$l$ CDW correlations persisting over long distances
in all three dimensions.

\begin{figure}[t!!!]
  \centering
  \includegraphics[width=\linewidth,clip=true]{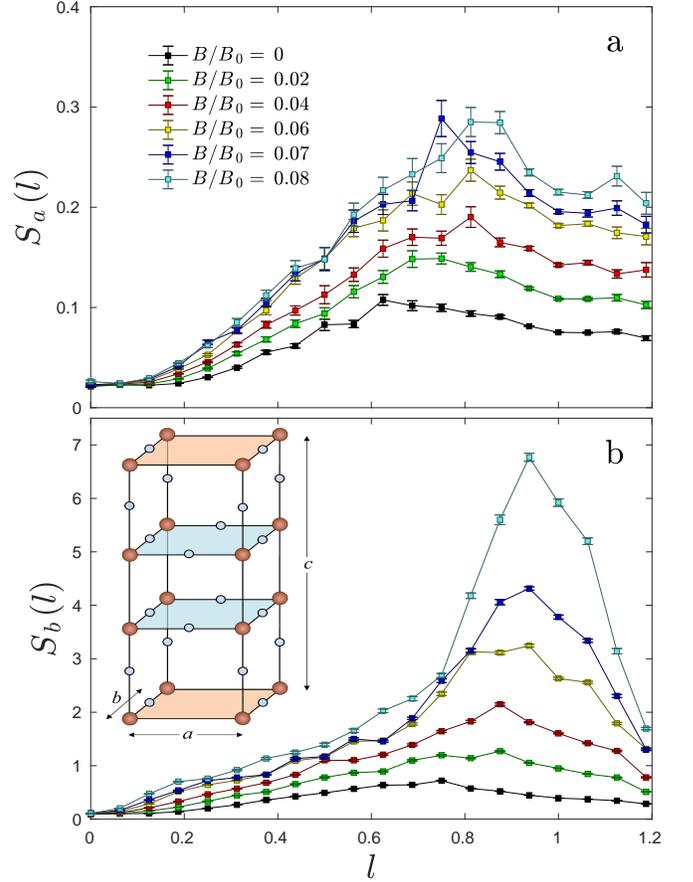}
  \caption{The CDW structure factor at the $a$ and $b$ incommensurate peaks as function of $c$-axis wave-vector, $l$, for $T=0.2\rhos$ and
  various magnetic fields. The inset depicts the YBCO unit cell. Only copper atoms (brown balls) and oxygen atoms
  (blue balls) are shown. The CuO$_2$ planes (light blue) host the SC and CDW orders.
  The doped oxygens go into the (orange) CuO$_x$ chain layers and are the main source of
  disorder.}
  \label{fig:Sab(l)}
\end{figure}

{\it The model.--} Our NLSM of YBCO consists of $N_c$ bilayers, see Fig. \ref{fig:Sab(l)},
hosting complex SC and CDW order parameters, $\psi_{\mu j}(\br)$ and $\Phi_{\mu j}^{a,b}(\br)$.
The latter describe density variations
$\delta\rho_{\mu j}=e^{i\bQ_a \cdot \br}\Phi_{\mu j}^a(\br)+e^{i\bQ_b \cdot \br}\Phi_{\mu j}^b(\br) +{\rm c.c.}$,
along the $a$ and $b$ directions with incommensurate wave-vectors $\bQ_{a,b}$.
Here, $j$ is the bilayer index and $\mu=0,1$ corresponds to the bottom (top) layer within a bilayer.
Focusing on $T<T_{ch}$ we assume the existence of some type of local order and the competition between
its components, as encapsulated by the constraints \cite{Hayward1,Nie,NLSM-layers}
\begin{equation}
\label{meq:constraint}
|\psi_{\mu j}|^2+|\bPhi_{\mu j}|^2=1,
\end{equation}
where ${\bm \Phi}_{\mu j}=(\Phi_{\mu j}^a,\Phi_{\mu j}^b)^T$. The Hamiltonian reads
\begin{eqnarray}
\label{meq:HV}
\nonumber
\!\!\!\!\!\!\!H&=&\sum_{\mu=0,1}\sum_{j=1}^{N_c}H_{\mu j}+\frac{\rhos}{2}\sum_{j=1}^{N_c}\int d^2r
\Bigg[\tU \bPhi^\dagger_{0j}\bPhi_{1j}\\
\nonumber
\!\!\!\!\!\!\!&&+U\bPhi^\dagger_{1j}\bPhi_{0j+1}
-\tJ\psi_{0j}^*\psi_{1j}-J\psi_{1j}^*\psi_{0j+1}\\
\!\!\!\!\!\!\!&&+{\bm V}^\dagger_j\left(\gamma \bPhi_{0j}
+ \bPhi_{1j}+\bPhi_{0j+1}+\gamma \bPhi_{1j+1}\right)+{\rm H.c.}\Bigg],
\end{eqnarray}
with the SC stiffness, $\rhos$, setting the overall energy scale.
We model the Coulomb interaction between CDW fields within a
bilayer by a local coupling $\tU$, and denote the intra-bilayer Josephson tunneling amplitude by $\tJ$.
The (weaker) Coulomb interaction and Josephson coupling between nearest-neighbor CuO$_2$ layers belonging to
adjacent unit cells are denoted by $U$ and $J$, respectively.
The disorder due to the doped oxygens on the chain layers couples via Coulomb interaction to the CDW fields.
We include its interaction with the neighboring bilayers, assuming that the coupling to the outer CuO$_2$ planes
is reduced by a factor $\gamma$ compared to the coupling to the inner CuO$_2$ planes. The disorder is described
by independent random Gaussian fields ${\bm V}_j=(V^1_j+iV^2_j,V^3_j+iV^4_j)^T$,
satisfying $\overline{V_j^\alpha}(\br)=0$ and
$\overline{V_j^\alpha(\br)V_{j'}^\beta(\br')}=V^2\delta_{\alpha\beta}\delta_{jj'}\delta(\br-\br')$,
with the overline signifying disorder averaging. Within a layer the physics is governed by
\begin{eqnarray}
\label{meq:slh}
\nonumber
\!\!\!\!\!\!\!H_{\mu j}&=&\frac{\rho_s}{2}\int\dr\Big[\left|(\bnabla+2ie\bA)\psi_{\mu j}\right|^2 +\lambda|\bnabla \bPhi_{\mu j}|^2\\
\!\!\!\!\!\!\!&&
+g|\bPhi_{\mu j}|^2+\Delta g |\Phi^a_{\mu j}|^2+\left(\tilde{\bm V}^\dagger_{\mu j}\bPhi_{\mu j}+{\rm H.c.} \right)\Big],
\end{eqnarray}
where $\lambda\rhos$ is the CDW stiffness and $g\rhos$ is the energy density penalty for CDW ordering.
The $\Delta g$ term reflects our assumption that the chain potential favors ordering
along the $b$-axis, either directly or via amplification by nematic interactions between the CDW components \cite{Nie-nematic,Hayward1,Nie}.
We consider the extreme type-II limit where the magnetic field, $B$, is uniform and points in the $c$-direction.
Therefore, we include only its orbital coupling to the SC order.
Finally, $\tilde {\bm V}_{\mu j}$ is the disorder potential on the CuO$_2$ layers, which we model by Gaussian random fields
with zero mean and
$\overline{\tilde V_{\mu j}^\alpha(\br)\tilde V_{\mu'j'}^\beta(\br')}=\tilde V^2\delta_{\alpha\beta}\delta_{\mu\mu'}\delta_{jj'}\delta(\br-\br')$.

\begin{figure}[t!!!]
  \centering
  \vspace{-0.08cm}
  \includegraphics[width=\linewidth,clip=true]{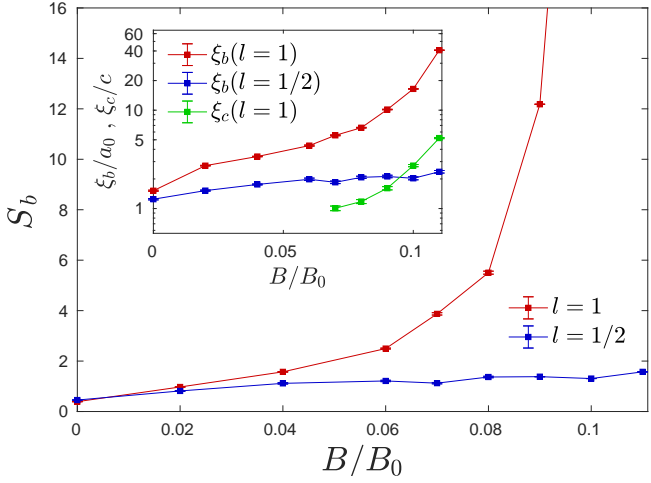}
  \caption{A magnetic field strongly enhances the $l=1$ CDW structure factor peak in the $b$-direction but only
  weakly affects the $l=1/2$ signal.
  Inset: in-plane, $\xi_b$, and out-of-plane, $\xi_c$, correlation lengths. Results are for $T=0.2\rhos$.}
  \label{fig:corrB}
\end{figure}
{\it Zero field.--} Our main interest lies in $k$-space (measured from $\bQ_{a,b}$) CDW correlations encapsulated by the matrix
\begin{eqnarray}
\label{meq:Gdef}
\nonumber
G^\alpha_{\mu\mu'}(\bq,l)&=&\frac{1}{2N_c A}\int d^2r d^2r'\sum_{jj'} e^{-i[\bq\cdot (\br-\br') + 2\pi (j-j')l]}\\
&&\times \overline{\langle\Phi^\alpha_{\mu j}(\br)\Phi^{*\alpha}_{\mu' j'}(\br')\rangle},
\end{eqnarray}
with $A$ the layer area. To make analytical progress we increase the number of independent components of $\Phi^{a,b}$ from two to large $N/2$,
assume $T\ll\rhos$, and use a saddle-point approximation \cite{suppmat}. For $B=0$, and $\gamma=\tilde V^2=0$,
$U\ll\tU$ we find (see Ref. \cite{suppmat} for the general result)
\begin{eqnarray}
\label{meq:GB0_00}
\nonumber
\!\!\!\!\!\!\!\!G^{\alpha}_{00}(0,l)&=&G^{\alpha}_{11}(0,l)
=\frac{T}{\rhos}\frac{\epsilon_\alpha}{\epsilon_\alpha^2-\epsilon_\perp^2(l)} \\
\!\!\!\!\!\!\!\!&&+\frac{V^2}{[\epsilon_\alpha+\epsilon_\perp(l)]^2}
+\frac{4V^2\epsilon_\alpha\tU\sin^2 \pi l}{[\epsilon_\alpha^2-\epsilon_\perp^2(l)]^2},\\
\nonumber
\label{meq:GB0_01}
\!\!\!\!\!\!\!\!G^{\alpha}_{01}(0,l)&=&G^{*\alpha}_{10}(0,l)
=-\frac{T}{\rhos}\frac{\tU}{\epsilon_\alpha^2-\epsilon_\perp^2(l)} \\
\!\!\!\!\!\!\!\!&&+V^2\frac{[(\epsilon_\alpha-\tU)\cos \pi l-i(\epsilon_\alpha+\tU)\sin\pi l]^2}
{[\epsilon_\alpha^2-\epsilon_\perp^2(l)]^2},
\end{eqnarray}
where $\epsilon_\alpha=g+\Delta g\delta_{\alpha a} +J+\tJ$, and
$\epsilon_\perp(l)=[U^2+\tU^2+2U\tU\cos 2\pi l]^{1/2}$.
Two ordering tendencies are apparent in Eqs. (\ref{meq:GB0_00}),(\ref{meq:GB0_01}).
While the temperature terms reach a maximum
at integer $l$, the disorder terms peak at half-integer $l$ as long as $\epsilon_\alpha>3U+\tU$, and dominate
the correlation matrix if $V^2>2UT/\rhos$, which is our case of interest. A small $\Delta g/g$ makes $\epsilon_a>\epsilon_b$
and introduces a slight tendency towards ordering along the $b$-axis.

\begin{figure}[t!!!]
  \centering
  \includegraphics[width=\linewidth,clip=true]{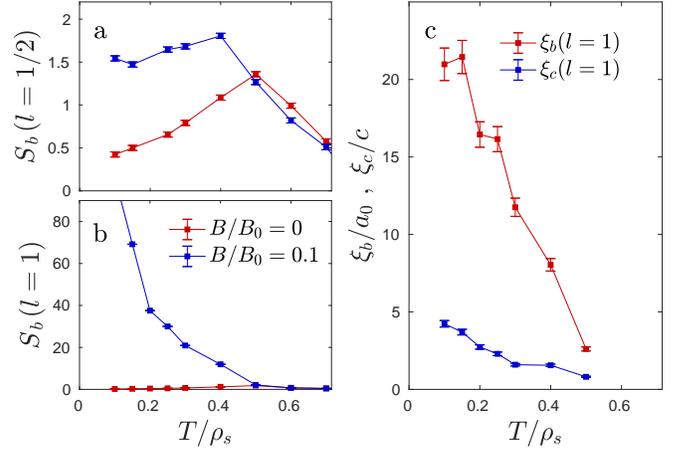}
  \caption{(a) Low-temperature high-field saturation of the $l=1/2$ CDW $b$-peak vs (b) increase of the $l=1$ signal.
  (c) In-plane and out-of-plane $l=1$ correlation lengths at $B=0.1B_0$.}
  \label{fig:corrT}
\end{figure}
To go beyond the limitations of the saddle-point approximation we used Monte-Carlo (MC) simulations of the
lattice version of Eqs. (\ref{meq:constraint})-(\ref{meq:slh}). The results are for a system of size
$64\times 64\times 32$ (16 bilayers), which is open in the $a$-direction, periodic in the $b$ and $c$-directions
and whose parameters are $ga_0^2=1.1$, $\Delta g a_0^2=0.1$, $\tJ a_0^2=0.15$, $Ja_0^2=0.015$, $\tU a_0^2=0.85$,
$Ua_0^2=0.12$, $V^2a_0^2=0.1$, $\gamma=0.15$, $\lambda=1$. Here, $a_0$ is
the in-plane lattice constant of the coarse grained model, which we assume is roughly the observed CDW wavelength,
\ie, about 3 Cu-Cu spacings. Each data point was averaged over 50-70 disorder realizations \cite{suppmat}.
In order to establish contact with the x-ray scattering experiments we obtained the CDW structure factor
$S_\alpha(\bq,l)$ by convolving $G$ with the measured CDW form factors \cite{Forgan-structure,suppmat}.
Fig. \ref{fig:Sab(l)} depicts the $l$ dependence of the structure factor at the in-plane peaks $S_{a,b}(l)=S_{a,b}(0,l)$.
We find that for $B=0$ both $S_a$ and $S_b$ exhibit a broad maximum centered around $l=0.6-0.7$, whose asymmetry is
largely due to the $l$ dependence of the form factors. The correlation lengths $\xi_{b,c}=1/\sigma_{b,c}$,
extracted from fits to $e^{-q^2/2\sigma^2}$, are both of order one lattice constant, see Fig. \ref{fig:corrB},
and the peak height reaches a maximum slightly above
$T_c\simeq 0.42\rhos$, see Fig. \ref{fig:corrT}(a), all qualitatively consistent with experiments.

{\it The effects of $B$.--} The suppression of superconductivity inside magnetically induced vortices facilitates CDW
nucleation there. At low $B$ these localized modes form a narrow band due to their small overlap and appear in tandem
to the more extended CDW states, which are already present at $B=0$ and produce the $l=1/2$ correlations. Their contribution
to $G$ is similar to Eqs. (\ref{meq:GB0_00}),(\ref{meq:GB0_01}) apart of two modifications that
determine its $B$ dependence \cite{suppmat}. First is an overall $B$-linear factor reflecting the number of vortices.
Secondly, $\epsilon_\alpha$ is given by the bottom of the vortex band, which drifts down with $B$ as CDW halos move
closer together. Consequently, the maximum of the disorder terms shifts towards integer $l$ and the entire $G$ increases
in magnitude. The effect, however, is very sensitive to the small anisotropy, $\Delta g$.
Our MC results, shown in Figs. \ref{fig:Sab(l)} and \ref{fig:corrB}, demonstrate that the CDW core regions orient predominantly
along the $b$ axis, causing $S_b$ to form a rapidly growing peak near $l=1$ for fields beyond $B_{3D}=0.06-0.07 B_0$.
Since $B_0=\phi_0/2\pi a_0^2\simeq 250 {\rm T}$, where $\phi_0=\pi/e$ is the flux quantum, this crossover scale
corresponds to 15-18T for the set of parameters used by us. At the same time Fig. \ref{fig:Sab(l)} shows that $S_a$
is only weakly modified by the presence of the field.

The agreement of the calculated high-field signal with the observed x-ray phenomenology \cite{Gerber3D,Chang3D,Jang3D}
extends beyond its unidirectionality and sharp $B$-dependence. Fig. \ref{fig:corrB} shows that the increase of
$S_b(l=1)$ is accompanied by a substantial growth of the $l=1$ correlation lengths. At $B=0.11B_0$ - the highest field
we could handle without significant finite size effects, $\xi_c$ extends over 5 $c$-axis lattice constants and
$\xi_b=40a_0$ is found within the planes. In contrast, the $l=1/2$ correlation lengths
change very little with $B$ and remain short. The dichotomy between the two types of correlations is also reflected
by their $T$-dependence, depicted in Fig. \ref{fig:corrT}. While the $T=0.5\rhos$ peak of the $l=1/2$ signal turns
into a low-temperature saturation in the presence of high magnetic fields \cite{Chang}, $S_b(l=1)$ and its
associated correlation lengths exhibit a rapid upturn below $T_{3D}=0.5\rhos$ at large $B$.

{\it Transition to long-range order.--} The following Imry-Ma argument \cite{Imry-Ma} shows that
in the absence of in-plane disorder the integer-$l$ CDW can become long-ranged. Consider a domain of linear size $L$ of such a CDW.
If the order is constant the interaction of the chain disorder with its neighboring CuO$_2$ planes cancels out.
However, if the order varies as $1/L$ along the $c$-axis the averaged squared interaction scales, in $d$ dimensions,
as $L^{d-2}$ and can lead to a typical energy gain of $L^{d/2-1}$. Since the elastic energy to create the
domains scales as $L^{d-2}$ their proliferation become 
favorable only in $d\leq 2$. Our large-$N$ analysis reflects this physics \cite{suppmat}.
For $\tilde V^2=0$ and $V^2\ll (r_0 U)^2 < U \ll \tU$ we find that $\xi_c$ diverges at
\begin{equation}
\label{meq:TCDWblu}
\frac{T_{CDW}}{\rhos}=\kappa r_0^2\sqrt{tU}-\frac{V^2}{2U},
\end{equation}
where $r_0$ is the vortex core radius, $t\sim r_0^{-2} e^{-b\sqrt{\phi_0/Br_0^2}}$ and $\kappa$,$b$ constants.
Hence, the clean system orders for any small magnetic field at low enough temperatures.
In the presence of chain disorder a transition occurs only above a critical field, which for $T=0$ is approximately
\begin{equation}
\label{meq:BCDWblu}
\frac{B_{CDW}r_0^2}{\phi_0}\approx\ln^{-2}\left[\kappa^2r_0^2 U\left(\frac{2U}{V^2}\right)^2\right].
\end{equation}
Similar expressions for the case of stronger disorder can be found in Ref. \cite{suppmat}.

\begin{figure}[t!!!]
  \centering
  \includegraphics[width=\linewidth,clip=true]{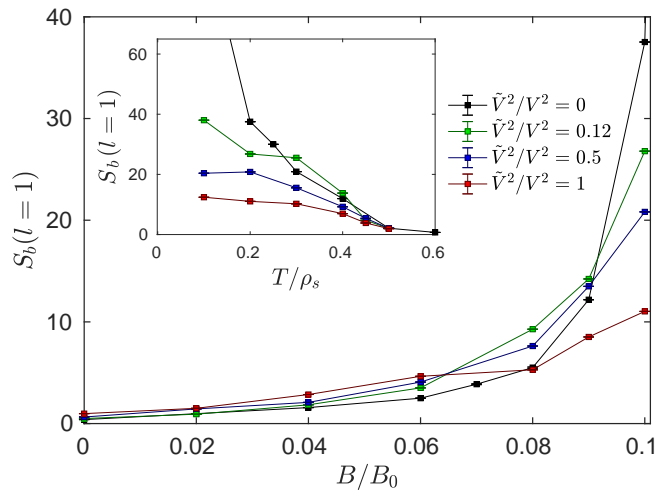}
  \caption{Suppression of the $l=1$ CDW $b$-peak by in-plane disorder, $\tilde{V}$, as revealed by its dependence on
  magnetic field (at $T=0.2\rhos$) and on temperature (at $B=0.1B_0$).}
  \label{fig:disorder}
\end{figure}

{\it The effects of in-plane disorder.--} Contrary to the chain disorder,
the in-plane disorder couples to each layer separately, leads to a typical energy gain which scales
as $L^{d/2}$, and thus prevents long-range order at $d\leq 4$ \cite{Imry-Ma}. Indeed, our
saddle-point equations \cite{suppmat} do not admit a diverging $\xi_c$ when $\tilde V^2>0$.
Fig. \ref{fig:disorder} shows that as $\tilde V^2/V^2$
approaches 1 the rapid increase of the $l=1$ correlations is averted. We therefore infer that in the physical systems
$\tilde V^2\ll V^2$.

{\it The sensitivity to $\Delta g$.--} It is difficult to ascertain the magnitude of the anisotropy in $g$. A proxy
might be the resistivity anisotropy which is roughly $\rho_a/\rho_b\approx 1.5$ in the relevant YBCO samples \cite{Ando02}.
Much larger values have been measured for the ratio of the Nernst coefficients \cite{Taileffer-rotational}. The presented
MC results are for $\Delta g/g=9\%$ and we have checked that deviations from a unidirectional $l=1$ signal commence
only around $\Delta g/g=3\%$.

{\it Discussion.--} 
Let us conclude by pointing out few consequences of our model. First, since the enhancement of CDW
by a magnetic field is driven by the suppression of superconductivity, one is led to infer the existence
of local SC order
as long as CDW correlations continue to increase with $B$. Surprisingly, in ortho-VIII YBCO \cite{Jang3D}
the $l=1$ scattering intensity and correlation volume grow up to $H=32$T, well in excess of the resistive critical
field $H_{c2}=24$T \cite{Hc2}. Hence, an interesting possibility arises that in this system local SC order
continues to exist long after global superconductivity is lost.

Secondly, like in YBCO the disorder due to doped oxygens in \HBCO~resides on planes (HgO) \cite{Campi} shared by
consecutive unit cells along the $c$-axis. The arguments presented above would then imply that in this single-layer
compound low-field CDW correlations should broadly peak near integer $l$. Zero-field measurements \cite{Tabis} found
CDW peaks at $l=1.12$ and $l=1.25$, but experimental constraints make it currently impossible to determine
whether these are the true maxima. In a magnetic field the interaction between CDW halos on neighboring
planes is expected to move the scattering peaks towards half-integer $l$. Since \HBCO~is tetragonal, with no
dopant order or signs of nematicity, the signal would likely remain bidirectional. On the other hand,
in the \LSCO~unit cell each of the two CuO$_2$ planes is separately affected, at least to first approximation,
by the Sr disorder on its adjacent LaO layers. Furthermore, consecutive CuO$_2$ planes are offset by half a
lattice constant and Coulomb interactions between next-nearest-neighbor planes dominate and lead to
half-integer-$l$ peaks at low fields \cite{Croft,Christensen}. We then expect a high field to strengthen
and sharpen the peaks without shifting their $l$.

\acknowledgments
This research was supported by the Israel Science Foundation (Grant No. 701/17) and by
the United States-Israel Binational Science Foundation (Grant No. 2014265).

\newpage

\onecolumngrid

\pagenumbering{gobble}

\setcounter{equation}{0}

\section*{Supplementary Material for "Dimensional Crossover of Charge-Density Wave Correlations in the Cuprates"}

\subsection{\fontsize{11}{13}\selectfont A. The model in the large-$N$ approximation}

With the aim of applying a saddle-point approximation to the model defined by Eqs. (1)-(3) of the main text, we enlarge
the number of components of $\Phi^{a,b}$ from 2 to $N/2$. The CDW order parameters are then described by the
real fields $n_{\mu j}^\alpha(\br)$, with $\alpha=1,\cdots, N/2$ corresponding to $\Phi^a$ and $\alpha=N/2+1,\cdots, N$
to $\Phi^b$. The Hamiltonian becomes
\begin{eqnarray}
  \label{eq:HV}
  \nonumber
  H&=&\frac{\rho_s}{2}\sum_{\mu=0,1}\sum_{j=1}^{N_c}\int\dr\left\{\left|(\bnabla+2ie\bA)\psi_{\mu j}\right|^2
  +\sum_{\alpha=1}^N\left[\lambda(\bnabla n^{\alpha }_{\mu j})^2+g(n_{\mu j}^{\alpha})^2
  +2\tilde V_{\mu j}^\alpha n_{\mu j}^{\alpha}\right]\right\}\\
  \nonumber
  &&  + \rhos\sum_{j=1}^{N_c}\int d^2r\,\left[ \sum_{\alpha=1}^N
  \left(\tU n^{\alpha }_{0j}n^{\alpha }_{1j}+U n^{\alpha }_{1j}n^{\alpha }_{0j+1}\right)
  -\frac{1}{2}\left(\tJ\psi_{0j}^*\psi_{1j}
  +J\psi_{1j}^*\psi_{0j+1}+{\rm c.c.}\right)\right]\\
  &&+ \rhos\sum_{j=1}^{N_c}\int d^2r\,\sum_{\alpha=1}^N
  V_j^{\alpha}\left(\gamma n^\alpha_{0j}+ n^\alpha_{1j}+n^{\alpha }_{0j+1}+\gamma n^\alpha_{1j+1}\right),
\end{eqnarray}
where we have set $\Delta g=0$ in order to slightly simplify the following analysis. We will adapt the results to the
general case where $g$ depends on $\alpha$ at the end of the calculation.
The constraints read
\begin{equation}
\label{eq:constraint}
|\psi_{\mu j}|^2+\sum_{\alpha=1}^N(n_{\mu j}^\alpha)^2=1.
\end{equation}
The partition function
\begin{equation}
  \label{eq:FV}
  Z=e^{-\beta F}=  \int\D\psi\D n^\alpha\,\prod_{\mu j}\delta\left[|\psi_{\mu j}|^2
    +\sum_\alpha (n_{\mu j}^\alpha)^2-1\right]e^{-S},
\end{equation}
is defined by the action
\begin{equation}
  \label{eq:HVsrcs}
  S = \beta H
   - \int d^2rd^2r'\sum_{\mu\mu' jj'}\sum_{\alpha\beta}K^{\alpha\beta} _{\mu\mu' jj'}(\br,\br')
  n_{\mu j}^{\alpha }(\br)n^{\beta }_{\mu' j'}(\br'),
\end{equation}
to which we have added a source term that yields the correlation function via
\begin{equation}
  \label{eq:G}
  G^{\alpha\beta}_{\mu\mu' jj'}(\br,\br')\equiv\overline{\braket{n^{\alpha}_{\mu j}(\br)n^{\beta}_{\mu' j'}(\br')}}
    =\left.-\frac{\delta\,\beta\overline{F}}{\delta K^{\alpha\beta}_{\mu\mu' jj'}(\br,\br')}
      \right|_{K=0}.
\end{equation}

To calculate the free energy averaged over realizations of disorder, $\overline{F}$, we employ
the replica method where we consider $m$ replicas of the original model, and use 
\begin{equation}
-\beta \overline{F}=\overline{\ln Z}=\lim_{m\rightarrow0}\frac{\overline{Z^m}-1}{m}\equiv\lim_{m\rightarrow0} \frac{e^{-\beta F(m)}-1}{m}.
\end{equation}
This implies that
\begin{equation}
\overline{F}=\lim_{m\rightarrow0} \frac{F(m)}{m}
\end{equation}
with $F(m)$ defined by
  \begin{eqnarray}
    \label{eq:Fm}
    \nonumber
    \!\!\!\!e^{-\beta F(m)}&=&\int\!\D\psi^{a}\D n^{\alpha a} \D V^\alpha \D \tilde V^\alpha
    \prod_{\mu j}\prod_{a=1}^m \delta\!\left[|\psi_{\mu j}^a|^2+\sum_\alpha(n_{\mu j}^{\alpha a})^2-1\right]\!
    e^{-\int d^2r \sum_{j \alpha} \left[ \frac{|V_j^\alpha|^2}{2V^2}+\sum_\mu \frac{|\tilde V_{\mu j}^\alpha|^2}{2\tilde V^2}\right]
    -\sum_{a=1}^m S[\psi^a,n^{\alpha a},K_{\alpha\beta}]}  \\
    \!\!\!\!&=&\int \!\D\psi^{a}\D n^{\alpha a} \D V^\alpha\D \tilde V^\alpha \D\bar\sigma^a
    e^{\int d^2r \left\{ i\sum_{\mu j a}\bar\sigma^a_{\mu j}\left[|\psi^a_{\mu j}|^2+\sum_\alpha(n_{\mu j}^{\alpha a})^2-1\right]
    -\sum_{j \alpha} \left[ \frac{|V_j^\alpha|^2}{2V^2}+\sum_\mu \frac{|\tilde V_{\mu j}^\alpha|^2}{2\tilde V^2}\right]\right\}
    -\sum_a S[\psi^a,n^{\alpha a},K_{\alpha\beta}]}.
  \end{eqnarray}
Integrating over $V_\alpha$, $\tilde V_\alpha$ and analytically continuing to $\bar\sigma_{\mu j}^a=i\frac{\beta\rhos}{2}\sigma_{\mu j}^a$
we find $e^{-\beta F(m)}=\int\D\psi^{a}\D n^{\alpha a}\D\sigma^a e^{-\tilde S(m)}$ given in terms of the effective action
\begin{eqnarray}
  \label{eq:Stilde}
  \nonumber
 \tilde S(m) &=& \frac{\beta\rhos}{2}\int d^2r \Bigg\{\sum_{j\mu}\sum_a \left[
      |(\bnabla+2ie\bA)\psi^a_{\mu j}|^2+\sigma^a_{\mu j}(|\psi^a_{\mu j}|^2-1)\right]
-\sum_j\sum_a\left(\tJ\psi_{0j}^{a*}\psi_{1j}^a +J\psi_{1j}^{a*}\psi_{0j+1}^a + {\rm c.c.}\right) \\
\nonumber
    &&+\sum_{ jj'}\sum_{\alpha}\sum_{ab} \left(\begin{array}{cc}
     n_{0 j}^{\alpha a} & n_{1 j}^{\alpha a}\end{array}\right)\left[\delta_{ab}\hat L^a_{ jj'}
      -\hat{M}_{jj'}\right] \left(\begin{array}{c} n_{0 j'}^{\alpha b} \\ n_{1 j'}^{\alpha b}\end{array}\right) \Bigg\}
     - \int d^2rd^2r'\sum_{\mu\mu' jj'}\sum_{\alpha}\sum_{a}K^{\alpha\beta}_{\mu\mu' jj'}(\br,\br')
  n^{\alpha a}_{\mu j}(\br)n^{\beta a}_{\mu' j'}(\br'),\\
\end{eqnarray}
where a hat denotes a $2\times 2$ matrix whose indices are $\mu,\mu'$, and
\begin{eqnarray}
\label{eq:Lmat}
\hat{L}_{ jj'}^a& =&\left( \begin{array}{cc}
\left[-\lambda\nabla^2+g+\sigma_{0 j}^a(\br)\right]\delta_{jj'} & \tU\delta_{j'j} +U\delta_{j'j-1} \\
\tU\delta_{j'j} +U\delta_{j'j+1}  & \left[-\lambda\nabla^2+g+\sigma_{1j}^a(\br)\right]\delta_{jj'} \end{array} \right),\\
\hat{M}_{ jj'}& =& \beta\rhos V^2 \left( \begin{array}{cc}
(1+\gamma^2+\upsilon^2)\delta_{jj'}+\gamma(\delta_{j' j-1}+\delta_{j' j+1}) & \delta_{j' j-1} +2\gamma\delta_{jj'}+\gamma^2\delta_{j' j+1} \\
\delta_{j' j+1} +2\gamma\delta_{jj'}+\gamma^2\delta_{j' j-1} & (1+\gamma^2+\upsilon^2)\delta_{jj'}+\gamma(\delta_{j' j-1}+\delta_{j' j+1})
\end{array} \right),
\end{eqnarray}
with $\upsilon^2={\tilde V}^2/V^2$. Next, we integrate over the CDW fields to obtain
$e^{-\beta F(m)} =\int\D\psi^{a}\D\sigma^ae^{-S(m)}$, where
\begin{eqnarray}
  \label{eq:Sm}
  S(m) & = &
  \frac{\beta\rhos}{2}\int d^2r\sum_{ja}\Bigg\{\sum_\mu\left[
    |(\bnabla+2ie\bA)\psi_{\mu j}^a|^2+\sigma_{\mu j}^a(|\psi_{\mu j}^a|^2-1)\right]
-\left(\tJ\psi_{0j}^{a*}\psi_{1j}^a+J\psi_{1j}^{a*}\psi_{0j+1}^a+{\rm c.c.}\right) \Bigg\}\nonumber \\
  & & +\frac{1}{2}\Tr\ln(G^{-1}-2K).
\end{eqnarray}
Here
\begin{equation}
  \label{eq:Ginv}
  (\hat G^{-1})^{\alpha\beta ab}_{ jj'}(\br,\br') = \beta\rhos\left[\delta_{ab}\hat L^a_{ jj'}
      -\hat M_{ jj'}\right]\delta_{\alpha\beta}\delta(\br-\br'),
\end{equation}
and
\begin{equation}
  \label{eq:K}
  K^{\alpha\beta ab }_{\mu\mu'jj'}(\br,\br')=\delta_{ab}K^{\alpha\beta}_{\mu\mu' jj'}(\br,\br').
\end{equation}

The integrals over $\psi^a_{\mu j}$ and $\sigma^a_{\mu j}$
are to be calculated using a saddle-point approximation, which is justified
in the limits $N\to\infty$ and $\beta\to\infty$, provided that the disorder is weak and satisfies
$V^2a^2={\cal O}(1/N)$, see Eq. (\ref{eq:psi0TB03}) below. Within this approximation, $\beta F(m)=S(m)$,
where $S(m)$ is evaluated using the saddle-point configurations, satisfying
\begin{eqnarray}
  \label{eq:SPV}
  \frac{\delta S(m)}{\delta\sigma_{\mu j}^a(\br)}
  &=& \frac{\beta\rhos}{2}\left[\sum_\alpha G^{\alpha\alpha aa}_{\mu\mu jj}(\br,\br)
    +(|\psi_{\mu j}^a|^2-1)\right]=0, \\
  \label{eq:SPVpsi}
  \frac{\delta S(m)}{\delta\psi_{\mu j}^{a*}(\br)}
  &=&\frac{\beta\rhos}{2}\left\{\left[-(\bnabla+2ie\bA)^2+\sigma^a_{\mu j}\right]\psi_{\mu j}^a
  -\tJ\psi_{\bar{\mu}j}^a -J\left(\delta_{\mu0}\psi_{1j-1}^a +
  \delta_{\mu1}\psi_{0j+1}^a\right)\right\}=0,
\end{eqnarray}
with $\bar{\mu}=1-\mu$. We have neglected ${\cal O}(K)$ terms in the above equations since they lead to an ${\cal O}(K^2)$
contribution to $S(m)$, which is irrelevant for the purpose of calculating $G$, as follows from Eq. (\ref{eq:G}).
Furthermore, since $G^{-1}$ is diagonal
in $\alpha,\beta$ and symmetric in $a,b$, $\br,\br'$, and under exchange of both $j,j'$ and $\mu,\mu'$ so is
$G^{\alpha\beta ab}_{\mu\mu' jj'} = \delta_{\alpha\beta}G_{\mu\mu'jj'}^{ab}$. Thus,
from Eqs. (\ref{eq:G}) and (\ref{eq:Sm}) one finds
\begin{equation}
  \label{eq:Glim}
  \hat G^{\alpha\beta}_{jj'}(\br,\br')\equiv \delta_{\alpha\beta}\hat G_{jj'}(\br,\br')=\delta_{\alpha\beta}\lim_{m\to 0}\frac{1}{m}
  \sum_a \hat G_{jj'}^{aa}(\br,\br').
\end{equation}

We will calculate the $2\times2$ correlation matrix $\hat G_{jj'}^{ab}$ by assuming a replica-symmetric
solution of the saddle-point equations, which is also independent of $j$ and $\mu$, \ie, $\psi_{\mu j}^a(\br)=\psi(\br)$
and $\sigma_{\mu j}^a(\br)=\sigma(\br)$. Under this assumption the operator $\hat L_{jj'}^a=\hat L_{jj'}$
is also replica symmetric, and $\hat G_{jj'}^{ab}$ is determined from
\begin{equation}
  \label{eq:GGinv}
\beta\rhos\sum_{cl}\int d^2\tilde r\left[\delta_{ac}\hat L_{jl} -\hat M_{jl}\right]\delta(\br-\tilde\br)
 \hat G_{lj'}^{cb}(\tilde\br,\br')=\hat I\delta_{jj'}\delta_{ab}\delta(\br-\br').
\end{equation}
Expanding $\hat G_{jj'}^{ab}$ in the eigen-basis of $L_0=-\lambda\nabla^2+g+\sigma(\br)$,
defined by $L_0\phi_s(\br)=\epsilon_s\phi_s(\br)$,
\begin{equation}
  \label{eq:Geigen}
  \hat G_{ jj'}^{ab}(\br,\br')=\frac{1}{N_c}\sum_{k_z}\sum_{st}\hat G_{ st}^{ab} e^{ik_z(j-j')c}\phi_s(\br)\phi_t^*(\br'),
\end{equation}
and plugging it into Eq. (\ref{eq:GGinv}) gives
\begin{equation}
\label{eq:Gdefeq}
\frac{\beta\rhos}{N_c}\sum_{c}\sum_{k_z}\sum_{st}\left[\delta_{ac}\hat {\tilde L}_s(k_z) -\hat{\tilde M}(k_z)\right]
\hat G_{ st}^{cb} e^{ik_z(j-j')c}\phi_s(\br)\phi_t^*(\br')=\hat I\delta_{jj'}\delta_{ab}\delta(\br-\br'),
\end{equation}
where $k_z$ is quantized in units of $2\pi/N_c c$, with $c$ the $c$-axis lattice constant, and
\begin{eqnarray}
\hat{\tilde L}_s(k_z)& =&\left( \begin{array}{cc}
\epsilon_s & \tU+U e^{-ik_z c}\\
\tU+U e^{ik_z c} & \epsilon_s  \end{array} \right),\\
\hat{\tilde M}(k_z)& =& \beta\rhos V^2 \left( \begin{array}{cc}
1+\gamma^2+\upsilon^2+2\gamma \cos(k_z c) & e^{ik_z c}\left(\gamma+e^{-ik_z c}\right)^2 \\
e^{-ik_z c}\left(\gamma+e^{ik_z c}\right)^2 & 1+\gamma^2+\upsilon^2+2\gamma \cos(k_z c) \end{array} \right).
\end{eqnarray}
Eq. (\ref{eq:Gdefeq}) is solved by
\begin{equation}
\label{eq:Gansatz}
\hat G^{ab}_{ st}=\delta_{st}\left[\hat A_s(k_z)\delta_{ab}+\hat B_s(k_z)\right],
\end{equation}
leading to the correlation matrix
\begin{equation}
\label{eq:Gsol}
\hat G_{ jj'}(\br,\br')=\frac{1}{N_c}\sum_{k_z}\sum_{s}\left[\hat A_s(k_z)+\hat B_s(k_z)\right]
e^{ik_z(j-j')c}\phi_s(\br)\phi_s^*(\br'),
\end{equation}
with
\begin{equation}
\label{eq:solA}
\hat A_s(k_z)=\frac{1}{\beta\rho_s}\hat {\tilde L}_s^{-1}(k_z)
=\frac{1}{\beta\rho_s}\frac{1}{\epsilon_s^2-\epsilon_\perp^2(k_z)}
\left(\begin{array}{cc}
\epsilon_s &  -\tU - U e^{-ik_z c} \\
-\tU - U e^{ik_z c}  & \epsilon_s
\end{array}\right),
\end{equation}
where
\begin{equation}
\label{eq:dispperp}
\epsilon_\perp(k_z)=\sqrt{U^2+\tU^2+2U\tU\cos(k_z c)},
\end{equation}
and
\begin{equation}
\label{eq:solB}
\hat B_s(k_z)=\frac{1}{\beta\rho_s}\left[\hat {\tilde L}_s(k_z)
-m\hat {\tilde M}(k_z)\right]^{-1}\hat {\tilde M}(k_z)\hat{\tilde L}_s^{-1}(k_z) \;\,
\overrightarrow{\scriptstyle{m\rightarrow 0}}\;\,\frac{1}{\beta\rho_s}\hat {\tilde L}_s^{-1}(k_z)
\hat {\tilde M}(k_z)\hat{\tilde L}_s^{-1}(k_z),
\end{equation}
whose components are
\begin{eqnarray}
\nonumber
B_{00}(k_z)&=&B_{11}(k_z)\\
\nonumber
&=&\frac{V^2}{\left[\epsilon_s^2-\epsilon_\perp^2(k_z)\right]^2}
\Big\{\left[1+2\gamma \cos(k_z c)+\gamma^2+\upsilon^2\right]\left[\epsilon_s^2+\epsilon_\perp^2(k_z)\right]\\
&&-2\epsilon_s\left[ U+2\gamma\tU+\left(2\gamma U+\tU+\gamma^2 \tU\right)
\cos(k_z c)+\gamma^2 U\cos(2k_z c)\right]\Big\},\\
\nonumber
B_{01}(k_z)&=&B_{10}^*(k_z)\\
&=&\frac{V^2e^{-ik_z c}}{\left[\epsilon_s^2-\epsilon_\perp^2(k_z)\right]^2}
\left\{\left[U+\gamma\tU-\epsilon_s+\left(\tU-\gamma\epsilon_s\right)e^{ik_z c}
+\gamma U e^{-ik_z c}\right]^2-2\upsilon^2\epsilon_s\left(U+\tU e^{ik_z c}\right)\right\}.
\end{eqnarray}

Let us comment on the changes incurred in the preceding analysis as a result of an $\alpha$ dependent $g$.
In such a case the operator $L_0$, appearing on the diagonal of Eqs. (\ref{eq:Lmat}), (\ref{eq:Ginv}), and (\ref{eq:GGinv}),
turns into $L_0^\alpha=-\lambda\nabla^2+g_\alpha+\sigma(\br)$. Its eigenfunctions $\phi_s$ are unchanged but the
spectrum, $\epsilon_s^\alpha$, is shifted and acquires $\alpha$ dependence. Consequently, so does $\hat{G}$, which now reads
\begin{equation}
\label{eq:Galpha}
\hat G_{ jj'}^\alpha(\br,\br')=\frac{1}{N_c}\sum_{k_z}\sum_{s}\left[\hat A^\alpha_s(k_z)+\hat B^\alpha_s(k_z)\right]
e^{ik_z(j-j')c}\phi_s(\br)\phi_s^*(\br'),
\end{equation}
where $\hat A^\alpha_s(k_z)$, and $\hat B^\alpha_s(k_z)$ are obtained from Eqs. (\ref{eq:solA}) and (\ref{eq:solB})
via the substitution $\epsilon_s\rightarrow\epsilon_s^\alpha$.

\subsection{\fontsize{11}{13}\selectfont B. The zero-field case}

In the absence of a magnetic field the saddle-point equations (\ref{eq:SPV},\ref{eq:SPVpsi}) possess a constant solution
$\psi_{\mu j}^a=\psi_0$, and $\sigma_{\mu j}^a=\tJ+J$. Consequently, the eigenfunctions of $L_0$ are plane waves
$\phi_s(\br)=\phi_{\bk}(\br)=\frac{1}{\sqrt{A}}e^{i{\bk}\cdot\br}$,
with eigenvalues $\epsilon^\alpha_s=\epsilon^\alpha_{\bk}=\lambda k^2+g_\alpha+\tJ+J$.
Hence the correlation matrix
\begin{equation}
\label{eq:G(q)}
\hat{G}^\alpha(\bq,q_z)=\frac{1}{N_c A}\int d^2r d^2r'\sum_{jj'} e^{-i[\bq\cdot (\br-\br') + q_z (j-j')c]}\hat{G}^\alpha_{jj'}(\br,\br'),
\end{equation}
takes the form
\begin{equation}
\label{eq:GqB0}
\hat{G}^\alpha(\bq,q_z)=\hat{A}^\alpha_{\bf q}(q_z)+\hat{B}^\alpha_{\bf q}(q_z).
\end{equation}

\begin{figure}[h]
\centering
  \includegraphics[width=9cm,clip=true]{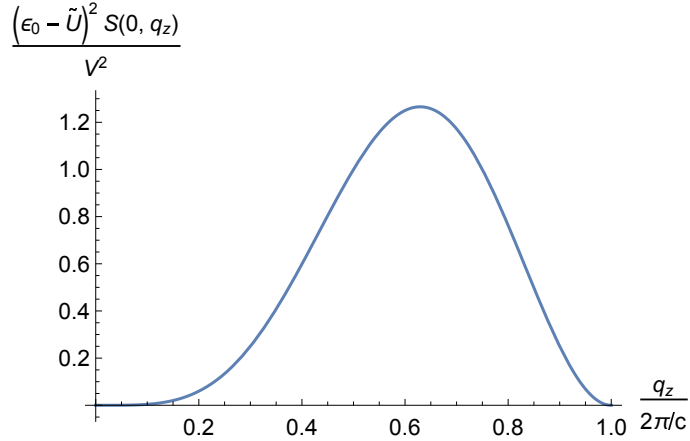}
\caption{$S(0,q_z)$ in the limit $\epsilon_0\gg\epsilon_0-\tU\gg U$, for $\gamma=\upsilon^2=T=0$.}
  \label{fig:S0}
\end{figure}

Neglecting the intra-cell form factor, with which we deal in Sec. D, and using the fact that the CuO$_2$ planes
within a bilayer are separated by approximately $c/3$ the contribution of the $\alpha$ component to the structure factor is given by
\begin{eqnarray}
\label{eq:Snaive}
\nonumber
S^\alpha(\bq,q_z)&=&\frac{1}{N_c A}\left\langle\left|\int d^2 r\sum_{\mu j}e^{-i[\bq\cdot \br +q_z(j+\mu/3)c]}
n^\alpha_{\mu j}(\br)\right|^2\right\rangle\\
&=&G_{00}^\alpha(\bq,q_z)+G_{11}^\alpha(\bq,q_z)+e^{iq_zc/3}G^\alpha_{01}(\bq,q_z)+e^{-iq_zc/3}G^\alpha_{10}(\bq,q_z).
\end{eqnarray}
The intensity at the incommensurate peak, for $\gamma=\upsilon^2=0$, is
\begin{equation}
\label{eq:G00}
S^\alpha(0,q_z)=2\frac{T}{\rhos}\frac{\epsilon^\alpha_0-\tU\cos\left(\frac{q_z c}{3}\right)-U\cos\left(\frac{2q_z c}{3}\right)}
{(\epsilon^\alpha_0)^2-\epsilon_\perp^2(q_z)}
+4V^2\frac{\left[(\epsilon^\alpha_0-U)\cos\left(\frac{q_z c}{3}\right)-\tU\cos\left(\frac{2q_z c}{3}\right)\right]^2}
{\left[(\epsilon^\alpha_0)^2-\epsilon_\perp^2(q_z)\right]^2},
\end{equation}
where $\epsilon^\alpha_0=g_\alpha+\tJ+J$. For $T/\rhos\ll V^2/(\epsilon^\alpha_0-\tU)$,
the intensity $S^\alpha(0,q_z)$ is dominated by the $V^2$ term and
has the form of an asymmetric peak. For $\epsilon^\alpha_0\gg\epsilon^\alpha_0-\tU\gg U$
it becomes $4V^2\sin^2(q_z c/2)\sin^2(q_z c/6)/(\epsilon^\alpha_0-\tU)^2$,
which reaches a maximum at $q_z=0.63 (2\pi/c)$, with FWHM=$0.42(2\pi/c)$, see Fig. \ref{fig:S0}.

Before moving on to include the effects of a magnetic field let us consider the saddle-point equation for $\sigma$, Eq. (\ref{eq:SPV}),
which for the case $g_\alpha=g$ reads
\begin{equation}
\label{eq:psi0TB01}
Nc\int \frac{d^2 k}{(2\pi)^2} \int_0^{2\pi/c} \frac{dk_z}{2\pi} [A_{00}(\bk,k_z)+B_{00}(\bk,k_z)]=1-|\psi_0|^2.
\end{equation}
For $\gamma=0$ this can be written as
\begin{eqnarray}
\label{eq:psi0TB02}
\nonumber
\frac{1-|\psi_0|^2}{N}&=&\left\{\frac{T}{\rhos}-V^2\left[(1+\upsilon^2)\frac{\partial}{\partial g}+\frac{\partial}{\partial U}\right]\right\}
\int \frac{d^2 k}{(2\pi)^2} \int_0^{2\pi} \frac{dk_z}{2\pi} \frac{\epsilon_\bk}{\epsilon_\bk^2-\epsilon_\perp^2(k_z/c)}\\
&=&\left\{\frac{T}{\rhos}-V^2\left[(1+\upsilon^2)\frac{\partial}{\partial g}+\frac{\partial}{\partial U}\right]\right\}
\int \frac{d^2 k}{(2\pi)^2}\frac{\epsilon_\bk}{\sqrt{(\epsilon_\bk^2-U^2-\tU^2)^2-(2U\tU)^2}},
\end{eqnarray}
where we assume $\epsilon_0>U+\tU$ to avoid divergence of the $k_z$ integral. The $\bk$ integral requires
ultraviolet regularization, which we achieve by restricting it to a disk of radius $\pi/a_0$. Defining the clean mean-field
transition temperature
\begin{eqnarray}
\label{eq:TMF0def}
\nonumber
\frac{1}{T_{MF}^0}&=&\frac{N}{\rhos}\int_0^{\pi/a_0} \frac{k d k}{2\pi}\frac{\epsilon_\bk}{\sqrt{(\epsilon_\bk^2-U^2-\tU^2)^2-(2U\tU)^2}}\\
&=&\left.\frac{N}{8\pi\lambda\rhos}\ln\left[\epsilon^2-U^2-\tU^2+\sqrt{(\epsilon^2-U^2-\tU^2)^2-(2U\tU)^2}\right]
\right|_{\epsilon=g+J+\tJ}^{\epsilon=\lambda(\pi/a_0)^2+g+J+\tJ},
\end{eqnarray}
we obtain
\begin{equation}
\label{eq:psi0TB03}
|\psi_0|^2=1-\frac{T}{T_{MF}^0}+\rhos V^2
\left[(1+\upsilon^2)\frac{\partial}{\partial g}+\frac{\partial}{\partial U}\right]\frac{1}{T_{MF}^0},
\end{equation}
where it can be checked that the last term is negative. Therefore, disorder reduces $|\psi_0|^2$ and $T_c$. It cannot be too strong
otherwise $\psi=0$, and one needs to take into account fluctuations in $\psi$.

\subsection{\fontsize{11}{13}\selectfont C. The system in a magnetic field}

In the presence of a magnetic field an Abrikosov lattice of vortices develops. Within each vortex core $\psi$ vanishes linearly with
the distance to the vortex center and $\sigma$ is reduced from its value $J+\tJ$ far from the vortex. From Bloch's theorem we know that
\begin{equation}
\label{eq:phiBloch}
\phi_s(\br)=\phi_{n\bk}(\br)=e^{i\bk\cdot \br}u_{n\bk}(\br),
\end{equation}
where $\bk$ is in the magnetic Brillouin zone (MBZ) and $u_{n\bk}(\br+\bR)=u_{n\bk}(\br)$ for any
vector $\bR$ in the Abrikosov lattice. Writing $\br=\tilde{\br}+\bR$, where $\tilde{\br}$ is in the
magnetic unit cell (m.u.c) and using the periodicity of $u_{n\bk}$ we obtain from Eqs. (\ref{eq:Geigen}), (\ref{eq:Gansatz}),
and (\ref{eq:G(q)}) that
\begin{equation}
\label{eq:GeigB}
  \hat G(\bq,q_z)=\frac{1}{A}\int_{m.u.c}d^2r d^2r'\sum_{n\bk}\left[\hat A_{n\bk}(q_z)+ \hat B_{n\bk}(q_z)\right]
   \sum_{\bR\bR'}e^{i(\bk-\bq)\cdot(\bR-\bR')}e^{i(\bk-\bq)\cdot(\br-\br')}u_{n\bk}(\br)u_{n\bk}^*(\br'),
\end{equation}
where in the reminder we specify to the case $g_\alpha=g$.
Let us decompose $\bq=\bQ+\bq'$, with $\bQ$ a reciprocal magnetic vector and $\bq'$ lying in the MBZ. Using
$\sum_{\bR}e^{i(\bk-\bq)\cdot \bR}=\sum_{\bR}e^{i(\bk-\bq')\cdot \bR}=N_v\delta_{\bk,\bq'}$, where $N_v$ is the
number of unit cells (vortices) in the Abrikosov lattice we obtain
\begin{equation}
\label{eq:GeigB1}
  \hat G(\bq,q_z)=\frac{N_v^2}{A}\sum_{n}\left[\hat A_{n\bq'}(q_z)+ \hat B_{n\bq'}(q_z)\right]
  \left|\int_{m.u.c}d^2r e^{-i\bQ\cdot\br}u_{n\bq'}(\br)\right|^2.
\end{equation}
For $R\gg r_0$, where $r_0$ is the core radius, we expect the scattering states of $\hat{L}$ to be close to plane waves
(somewhat reduced inside the cores), with a spectrum that is close to the free dispersion. The index $n$ for these states
is the band index originating from folding the free dispersion onto the MBZ, and is thus given by a reciprocal wavevector
$\bK$, \ie,
\begin{eqnarray}
\label{eq:scaterstate}
u_{n\bq'}(\br)&=&u_{\bK\bq'}(\br)\simeq \frac{1}{\sqrt{A}}e^{i\bK\cdot \br},\\
\epsilon_{n\bq'}&=&\epsilon_{\bK\bq'}\simeq \epsilon_0+\lambda q'^2.
\end{eqnarray}
As a result, $\int_{m.u.c}d^2r e^{-i\bQ\cdot\br}u_{n\bq'}(\br)\simeq\frac{R^2}{\sqrt{A}}\delta_{\bK,\bQ}$ leading to
\begin{equation}
\label{eq:scattint}
 \left|\int_{m.u.c}d^2r e^{-i\bQ\cdot\br}u_{n\bq'}(\br)\right|^2=A_1\frac{A}{N_v^2}\delta_{\bK,\bQ},
\end{equation}
where $A_1<1$ is a numerical factor expressing the suppression of the scattered waves inside the cores. Thus,
the contribution of the scattering state to the correlation matrix is given approximately by
\begin{equation}
\label{eq:GqB0scatter}
\hat{G}_{scatt}(\bq,q_z)=A_1\left[\hat{A}_{\bf q}(q_z)+\hat{B}_{\bf q}(q_z)\right],
\end{equation}
where we have used the fact that $\hat{A}_{\bQ \bq'}(q_z)=\hat{A}_{\bf q}(q_z)$, and similarly for $\hat{B}$.
\newpage

The reduction in $\sigma$ within the vortex core acts as an attractive potential for the CDW fields. Consequently,
one expects that in the presence of a vortex the spectrum of $\hat{L}$ contains in addition to the scattering states
also a discrete set of bound states within the core. Our numerical solution of the saddle-point equations confirms
that there is a single (normalized) bound state, $\varphi_v(\br)$, that decays at large distances as $\varphi_v(\br)\sim \exp(-r/r_0)$.
In the presence of a dilute Abrikosov lattice of vortices, \ie, $R\gg r_0$, the small overlap between bound states
in neighboring cores leads to the formation of a tight-binding band $\phi_{v\bk}(\br)$. For a square Abrikosov lattice
(assuming a triangular lattice yields similar results with modified numerical constants), $R=\sqrt{\phi_0/B}$ and
\begin{equation}
\label{eq:uv}
u_{v\bk}(\br)=\frac{1}{\sqrt{N_v}}\sum_\bR e^{i\bk\cdot(\bR-\br)}\varphi_v(\br-\bR),
\end{equation}
with
\begin{equation}
\label{eq:epsilonv}
\epsilon_{v\bk}=\epsilon_v-2t\left[\cos(k_x R)+\cos(k_y R)\right],
\end{equation}
where the bound state eigenvalue, $\epsilon_v$, includes the shift due to the change in the effective core potential
induced by the other vortices, and where the overlap integral scales as $t\sim r_0^{-2} e^{-b\sqrt{\phi_0/Br_0^2}}$
with a constant $b$. Assuming an exponential bound state
$\varphi_v(r)=(2\pi r_0^2)^{-1/2}e^{-r/2r_0}$ we find for $qr_0\ll 1$
\begin{equation}
\label{eq:uvint2}
\left|\int_{m.u.c} d^2r e^{-i\bQ\cdot\br} u_{v\bq'}(\br)\right|^2=32\pi\frac{r_0^2}{N_v},
\end{equation}
which leads, together with $\epsilon_v(\bq')=\epsilon_v(\bq'+\bQ)=\epsilon_v(\bq)$ and Eq. (\ref{eq:GeigB1}),
to the vortex part of the correlation matrix
\begin{equation}
\label{eq:GqB0vortex}
\hat{G}_v(\bq,q_z)=32\pi\frac{r_0^2}{R^2}\left[\hat{A}_{v\bf q}(q_z)+\hat{B}_{v\bf q}(q_z)\right].
\end{equation}
Here, $\hat{A}_{v\bf q}(q_z)$ and $\hat{A}_{v\bf q}(q_z)$ are obtained by substituting $\epsilon_s=\epsilon_{v\bq}$
in Eqs. (\ref{eq:solA}) and (\ref{eq:solB}).

\subsection{The transition to long-range CDW order}

The expressions for the elements of $\hat{G}_v(0,q_z)$, which determine the vortex contribution to the scattering peak intensity,
all share the denominator $(\epsilon_v-4t)^2-U^2-\tU^2-2U\tU\cos(q_zc)$, or its square. For small $q_zc$, it can be written as
$U\tU c^2(\xi_c^{-2}+q_z^2)$, where the inverse correlation length is
\begin{equation}
\label{eq:chi2def}
\xi_c^{-2}=\frac{1}{c^2}\frac{(U+\tU)^2}{U\tU}\left[\left(\frac{\epsilon_v-4t}{U+\tU}\right)^2-1\right].
\end{equation}
We are interested in calculating the conditions under which $\xi_c$ diverges, signaling long-range CDW order with $q_z=0$.
To this end, we return to the saddle-point equation for $\sigma$,
\begin{equation}
\label{eq:spsigmaB}
\frac{N}{N_c}\sum_{k_z}\sum_s\left[\hat{A}_s(k_z)+\hat{B}_s(k_z)\right]_{00}
\left|\phi_s(\br)\right|^2=1-\left|\psi(\br)\right|^2.
\end{equation}
The contribution of the scattering states to the left hand side equals $1-|\psi_0|^2-|\delta\psi(\br)|^2$, where $\delta\psi$ is
appreciable only within the vortex cores. Thus, we conclude that the contribution of the vortex band is
\begin{equation}
\label{eq:spsigmavortex}
\int_{MBZ}\frac{d^2k}{(2\pi)^2}\int_0^{2\pi} \frac{dk_z}{2\pi}\left[\hat{A}_v(\bk,k_z/c)+\hat{B}_v(\bk,k_z/c)\right]_{00}
\left|\phi_{v\bk}(\br)\right|^2=\frac{|\psi_0|^2+|\delta\psi(\br)|^2-\left|\psi(\br)\right|^2}{NA}.
\end{equation}
Next, we integrate the equation over the plane using $\int d^2r |\phi_{v\bk}(\br)|^2=1$ and
$\int d^2r [|\psi_0|^2+|\delta\psi(\br)|^2-\left|\psi(\br)\right|^2] = N_v{\cal C}r_0^2|\psi_0|^2$,
with ${\cal C}$ a constant. This leads to
\begin{equation}
\label{eq:spsigmavortex2}
\int_{MBZ}\frac{d^2k}{(2\pi)^2}\int_0^{2\pi} \frac{dk_z}{2\pi}\left[\hat{A}_v(\bk,k_z/c)+\hat{B}_v(\bk,k_z/c)\right]_{00}
=\frac{{\cal C}r_0^2|\psi_0|^2}{NR^2}.
\end{equation}
For $\gamma=0$, this can be expressed as
\begin{eqnarray}
\label{eq:spsigmavortex3}
\nonumber
\frac{{\cal C}r_0^2|\psi_0|^2}{NR^2}&=&\left\{\frac{T}{\rhos}-V^2\left[(1+\upsilon^2)\frac{\partial}{\partial \epsilon_v}
+\frac{\partial}{\partial U}\right]\right\}
\int_{MBZ} \frac{d^2 k}{(2\pi)^2} \int_0^{2\pi} \frac{dk_z}{2\pi} \frac{\epsilon_{v\bk}}{\epsilon_{v\bk}^2-\epsilon_\perp^2(k_z/c)}\\
&=&\left\{\frac{T}{\rhos}-V^2\left[(1+\upsilon^2)\frac{\partial}{\partial \epsilon_v}+\frac{\partial}{\partial U}\right]\right\}
\int_{MBZ} \frac{d^2 k}{(2\pi)^2}\frac{\epsilon_{v\bk}}{\sqrt{(\epsilon_{v\bk}^2-U^2-\tU^2)^2-(2U\tU)^2}}.
\end{eqnarray}
We evaluate the remaining integral in the limit $\epsilon\equiv(\epsilon_v-4t)/(U+\tU)\rightarrow 1$,
corresponding to a transition to long-range CDW order with integer $q_zc/2\pi$, see Eq. (\ref{eq:chi2def}).
In this limit the integral is dominated by
small momenta and we expand $\epsilon_{v\bk}$ to order $k^2$. Furthermore, we approximate the MBZ by a disk of radius $\pi/R$.
We have checked that these approximations lead only to an overall factor of order 1 relative to an exact numerical evaluation of the integral.

Eq. (\ref{eq:spsigmavortex3}) admits a solution with $\epsilon=1$ provided $\upsilon^2=0$. The $\upsilon^2$ term, on the
other hand, diverges as $1/\sqrt{\epsilon-1}$, precludes such a solution and smears the transition into a crossover. Concentrating
on the case $\upsilon^2=0$, and assuming $U\ll\tU$ along with $t\ll\tU$, one finds from Eq. (\ref{eq:spsigmavortex3})
that long-range order onsets when
\begin{equation}
\label{eq:condsmallb}
\frac{\pi}{\sqrt{2}}\kappa r_0^2t=\frac{T}{\rhos}\ln\left[\sqrt{\frac{\pi^2t}{2U}}+\sqrt{1+\frac{\pi^2t}{2U}}\right]
+\frac{V^2}{2U}\frac{1}{\sqrt{1+2U/\pi^2t}},
\end{equation}
where
\begin{equation}
\label{eq:kappadef}
\kappa=\frac{2^{5/2}{\cal C}|\psi_0|^2}{N}.
\end{equation}

For $t\ll U$ Eq. (\ref{eq:condsmallb}) implies an ordering temperature
\begin{equation}
\label{TCDWblu}
\frac{T_{CDW}}{\rhos}=\kappa r_0^2\sqrt{tU}-\frac{V^2}{2U}.
\end{equation}
The critical field is given by
\begin{equation}
\label{tCDWblu}
r_0^2t_{CDW}=\frac{1}{\kappa^2r_0^2 U}\left[\frac{T}{\rhos}+\frac{V^2}{2U}\right]^2,
\end{equation}
or more explicitly,
\begin{equation}
\label{BCDWblu}
\frac{B_{CDW}r_0^2}{\phi_0}\approx\ln^{-2}\left[\kappa^2r_0^2 U\left(\frac{T}{\rhos}+\frac{V^2}{2U}\right)^{-2}\right].
\end{equation}
Hence, we conclude that the clean system orders for any small magnetic field at low enough temperatures.
In the presence of disorder a transition occurs, \ie, $T_{CDW}\geq 0$, only if the field is sufficiently
strong $r_0^2t\geq(V^2/2U)^2/\kappa^2r_0^2 U$. This condition is compatible with the assumption $t\ll U$
only for weak disorder satisfying $V^2\ll (r_0 U)^2$. To maintain the condition of well separated vortices, \ie,
$r_0^2 t\ll 1$, it suffices to require in addition $r_0^2U<1$.

For $U\ll t$ the ordering temperature is
\begin{equation}
\label{eq:TCDWulb}
\nonumber
\frac{T_{CDW}}{\rhos}=\left[\sqrt{2\pi^2}\kappa r_0^2 t-\frac{V^2}{U}\right]\ln^{-1}\left[\frac{2\pi^2t}{U}\right].
\end{equation}
At low temperatures, such that $T/\rhos\ll(V^2/U)/\ln(t/U)$ the critical field is given by
\begin{equation}
\label{tCDWulb1}
r_0^2t_{CDW}=\frac{1}{\sqrt{2\pi^2}\kappa}\frac{V^2}{U},
\end{equation}
or
\begin{equation}
\label{BCDWulb1}
\frac{B_{CDW}r_0^2}{\phi_0}\approx \ln^{-2}\left[\sqrt{2\pi^2}\kappa\frac{U}{V^2}\right].
\end{equation}
This result fulfills $U\ll t$ and $r_0^2t\ll 1$ provided the disorder satisfies $U\gg V^2\gg (r_0 U)^2$. At higher temperatures satisfying
$T/\rhos\gg(V^2/U)/\ln(t/U)$
\begin{equation}
\label{tCDWulb2}
r_0^2t_{CDW}=\frac{1}{\sqrt{2\pi^2}\kappa}\frac{T}{\rhos}\ln\left[\frac{\sqrt{2\pi^2}}{\kappa r_0^2 U}\frac{T}{\rhos}\right]+r_0^2\Delta t,
\end{equation}
where $\Delta t/t\rightarrow 0$ as $r_0^2U\rightarrow 0$.

\subsection{\fontsize{11}{13}\selectfont D. Making contact with the x-ray scattering experiments}

To relate our model to x-ray measurements of the cuprates we assume that each lattice point in the model,
$\bR=(\br,j)=(m,n,j)*(3a,3b,c)$, corresponds to a supercell comprising of $3\times 3$ \YBCO$\,$ unit cells, containing approximately
one CDW oscillation in each direction, with $n_{\mu=0,1}$ corresponding to the amplitude of the CDW in the lower (upper)
half of the supercell. The x-ray scattering intensity is proportional to the structure factor
\begin{equation}
\label{eq:xraysf}
S(\bQ)=\left|\sum_\bR\sum_i f_i(\bQ)e^{i\bQ\cdot[\bR+\br_i+\bu_i(\bR)]}\right|^2,
\end{equation}
where $\br_i$ are the equilibrium positions of the ions in the supercell, $\bu_i$ are the deviations from these
positions and $f_i$ are ionic structure factors. Assuming small deviations we find
\begin{equation}
\label{eq:xraysfapp}
S(\bQ)\approx\left|\sum_\bR\sum_i f_i(\bQ)\left[1+i\bQ\cdot \bu_i(\bR)\right]e^{i\bQ\cdot(\bR+\br_i)}\right|^2.
\end{equation}

Consider, for example, the ordered CDW state along the $b$-axis, whose in-plane wave-vector we approximate by $(0,2\pi/3b)$.
The deviations in the lower ($\mu=0$) and upper ($\mu=1$) halves of the supercell [in which we include half of the
central Y ion and half of the bottom (top) chain layer] take the form
\begin{equation}
\label{eq:loweru}
\bu_{\mu,i}=(-1)^\mu u_i^b\cos\left(\frac{2\pi}{3}b_i\right)\hat{b}-u_i^c\sin\left(\frac{2\pi}{3}b_i\right)\hat{c},
\end{equation}
where $b_i$ is the position along the $b$-axis (in units of $b$) of the $i$th ion. The $u_i^{b,c}$ are the
displacement amplitudes of the group of ions to which the $i$th ion belong, \ie, Y, Ba, etc.
Next, we assume that in the thermally fluctuating state the deviations are proportional to the CDW amplitudes
of the NLSM
\begin{eqnarray}
\label{eq:lowerufluct}
\nonumber
\bu_{\mu,i}&=&{\rm Re}\left\{(n_\mu^3+in_\mu^4)\left[u_i^b e^{i 2\pi b_i/3}\hat{b}+(-1)^\mu u_i^ce^{i(2\pi b_i/3+\pi/2)}\hat{c}\right]\right\}\\
&=&u_i^b\left[n_\mu^3\cos\left(\frac{2\pi}{3}b_i\right)-n_\mu^4\sin\left(\frac{2\pi}{3}b_i\right)\right]\hat{b}
-(-1)^\mu u_i^c\left[n_\mu^3\sin\left(\frac{2\pi}{3}b_i\right)+n_\mu^4\cos\left(\frac{2\pi}{3}b_i\right)\right]\hat{c}.
\end{eqnarray}
Here we chose the sign such that full anti-phase between $n_0$ and $n_1$ reproduces the ordered CDW state whose
sign changes from one plane to the next within a CuO$_2$ bilayer.

Averaging over thermal fluctuations and using $\langle n(\bR)\rangle =0$ one obtains
\begin{equation}
\label{eq:xraysfappave}
S(\bQ)\approx \sum_{\bR,\bR'}\sum_{\mu\mu'i i'} f_i(\bQ)f_{i'}^*(\bQ)\left[1+\langle \bQ\cdot \bu_{\mu,i}(\bR)\bQ\cdot
\bu_{\mu',i'}(\bR') \rangle\right]e^{i\bQ\cdot(\bR-\bR'+\br_{\mu,i}-\br_{\mu',i'})}.
\end{equation}
For $\bQ=(0,k,l)*(2\pi/a,2\pi/b,2\pi/c)$ the part which gives a CDW scattering peak is
\begin{equation}
\label{eq:sfCDW}
S^b_{CDW}(k,l)=\sum_{\bR,\bR'}\sum_{\mu\mu'i i'} f_i f_{i'}^* \langle \bQ\cdot \bu_{\mu,i}(\bR)\bQ\cdot
\bu_{\mu',i'}(\bR') \rangle e^{i2\pi l(j-j')}e^{i[2\pi k(b_i-b_i')+2\pi l (c_{\mu,i}-c_{\mu',i'})]}.
\end{equation}
Using the fact that $\langle n^\alpha n^\beta\rangle \propto \delta_{\alpha\beta}$ and independent of $\alpha$ we have
\begin{equation}
\label{eq:sfCDW1}
S^b_{CDW}(k,l)=\sum_{\bR,\bR'}\sum_{\mu\mu'} \langle n_\mu(\bR)n_{\mu'}(\bR')\rangle F^b_{\mu\mu'}(k,l)e^{i2\pi l(j-j')},
\end{equation}
where the form factor is given by
\begin{eqnarray}
\label{eq:Fl}
\nonumber
F_{\mu\mu'}^b(k,l)&=&(2\pi)^2\sum_{i i'}f_i f_{i'}^*\Bigg\{\left[\left(\frac{k}{b}\right)^2 u_i^b u_{i'}^b
+\left(\frac{l}{c}\right)^2(-1)^{\mu+\mu'}u_i^c u_{i'}^c\right]\cos\left[\frac{2\pi}{3}(b_i-b_i')\right]\\
&&+ \frac{kl}{bc}\left[(-1)^{\mu'}u_i^b u_{i'}^c
-(-1)^{\mu}u_i^c u_{i'}^b\right]\sin\left[\frac{2\pi}{3}(b_i-b_i')\right]\Bigg\}
e^{i[2\pi k(b_i-b_{i'})+2\pi l (c_{\mu,i}-c_{\mu',i'})]}.
\end{eqnarray}

\newpage
\subsection{\fontsize{11}{13}\selectfont E. Disorder averaging of the Monte-Carlo data}

\begin{figure}[h]
\centering
  \includegraphics[width=10.5cm,clip=true]{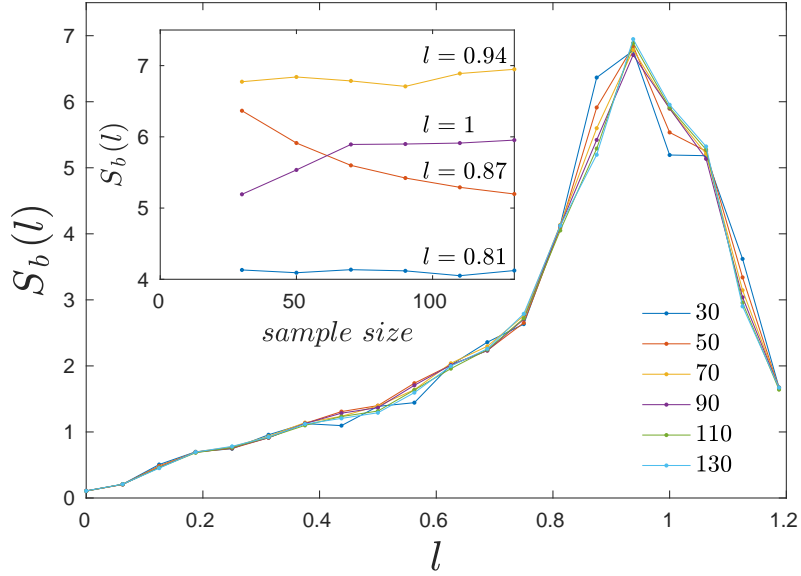}
\caption{The CDW structure factor at the $b$-peak as function of the $c$-axis wave-vector, $l$, for $T=0.2\rhos$
and $B=0.08B_0$. The data was averaged over an increasing number of disorder realizations, as indicated. The inset depicts
the evolution of the points around the maximum with the disorder sample size.}
  \label{fig:S1}
\end{figure}

We have found that for our relatively large system of 64$\times$64$\times$32 sites disorder averaging of the
Monte-Carlo results converged rather quickly. Among the various quantities which we have calculated, the
$l$ dependence of the CDW structure factor turned out to be the slowest to converge. Nevertheless, as shown
in Fig. \ref{fig:S1}, apart from occasional isolated points it becomes practically independent of the disorder
sample size, once more than 70 disorder realizations are included in the averaging.

\end{document}